\documentclass[10pt,a4paper]{article}
\usepackage{color}
\usepackage{graphicx}
\usepackage{authblk}
\usepackage{geometry}
\def\half{\frac{1}{2}}
\def\quarter{\frac{1}{4}}

\begin{document}

\title{\bf Scattering states and bound states of exponential potentials}

\author{Zafar Ahmed\footnote{{\tt zahmed@barc.gov.in}} \  and H.~F.~Jones\footnote{{\tt h.f.jones@imperial.ac.uk}}}

\affil{$^*$\emph{Nuclear Physics Division, Bhabha Atomic Research Centre, Mumbai 400\ 085, India}\\
$^\dag$\emph{Physics Department, Imperial College, London SW7 2BZ, UK}}

\date{February 11, 2021}

\maketitle

\begin{abstract}
We explore the relationships between scattering states and bound states of different non-analytic segments (depending on $|x|$) of the exponential potential, and elucidate the status of the special scattering states found in an earlier publication by Ahmed et al. A similar analysis can be made of non-analytic segments of power potentials such as $x^3$.
\end{abstract}

\section{Introduction}
In a recent paper\cite{Ahmed1} the non-analytic exponential potential well $V=V_0(e^{2|x|/a}-1)$ and its negative, the bottomless potential $V=V_0(1-e^{2|x|/a})$, were studied. These are denoted by $V_1$ and $V_4$ in what follows. They are potentials where the wave-functions are Bessel functions of various kinds. For the well, which has conventional bound states, the energy eigenvalues are the solutions of transcendental equations for the zeros of Bessel functions or their derivatives. For the bottomless potential the reflection and transmission coefficients were found, and their poles when the sign of $V_0$ was reversed were shown to give the bound-state energies of the potential well. In a second paper\cite{Ahmed2} interest was focussed on the bottomless potential $V_4$, and certain discrete states were identified within the continuum of scattering states. At the time, the precise significance of these states was not clear, but it was noticed in numerical studies\cite{BRath} that their energies corresponded to the bound-state energies of another related potential  ($V_5$ below).

In this paper we establish the connection between these special scattering states of $V_4$ and the bound states of $V_5$. Further, at these energies there occur special solutions for the bottomless analytic potential $V_6\equiv V_0(1-e^{-2x/a})$, which agrees with $V_4$ on the left ($x<0$) and $V_5$ on the right. Similar relationships link special scattering states of $V_2\equiv V_0(e^{-2|x|/a}-1)$ to the bound states of $V_1$ and special solutions of the analytic potential $V_3 \equiv V_0(e^{-2x/a}-1)$. Finally the relation between poles of the scattering amplitudes of $V_4$ and the bound states of $V_1$ established in Ref.~\cite{Ahmed1} is extended to a similar connection between the poles of the scattering amplitudes of $V_2$ and the bound states of $V_5$.

The last section is devoted to a discussion of the nature of the special scattering states of the bottomless potentials. They are normalizable, so in a sense can be regarded as unusual bound states, but their momentum expectation values diverge. A similar analysis can be applied to the cubic potential and its non-analytic pieces, but in this case higher moments of both $x$ and $p$ diverge.

\section{The exponential potential and its variants}
\begin{center}
\begin{figure}[h!]
\resizebox{!}{3cm}{\includegraphics{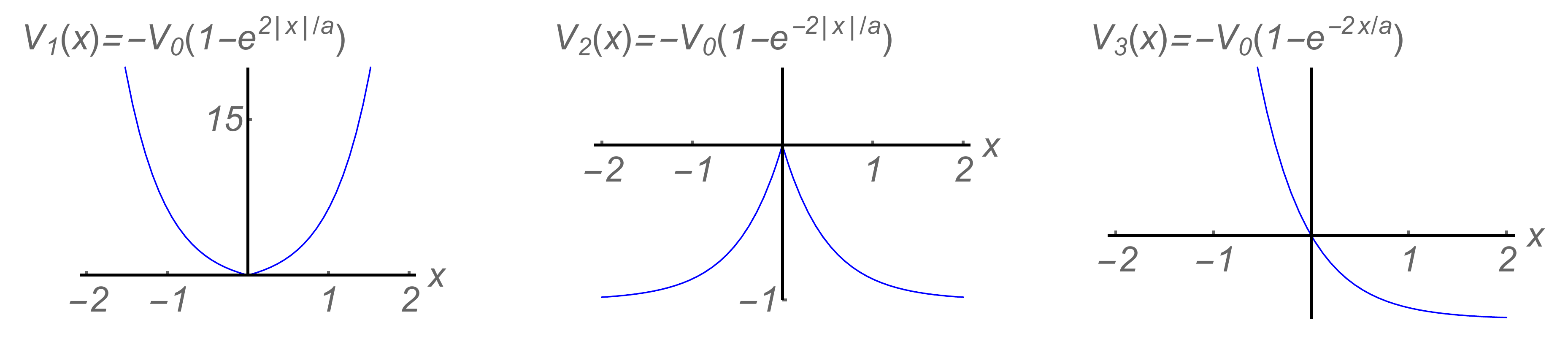}}\vspace{1cm}
\resizebox{!}{3cm}{\includegraphics{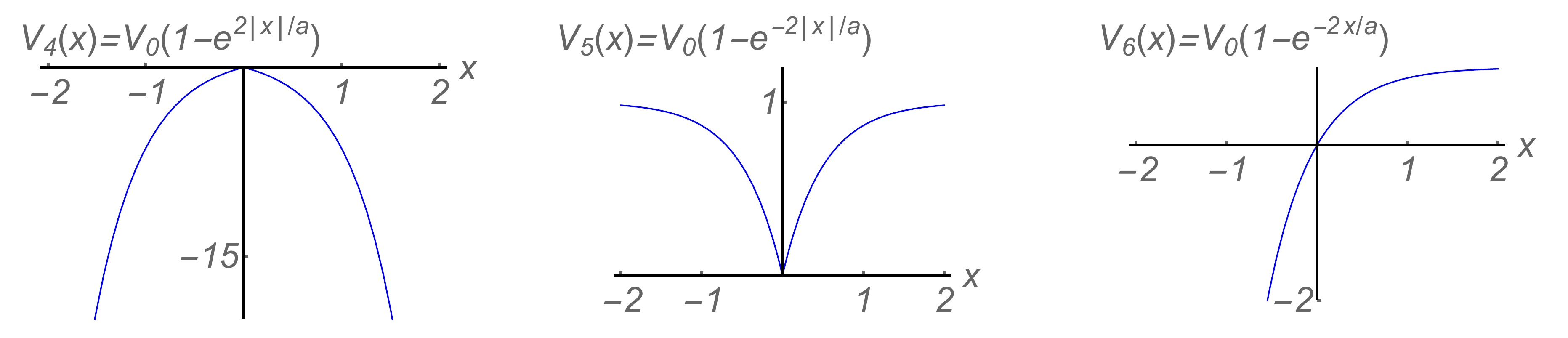}}
\caption{The various potentials associated with those of Ref.~\cite{Ahmed1} and \cite{Ahmed2}. For ease of illustration we have taken $V_0=a=1$.}
\end{figure}
\end{center}
\vspace{-1cm}
We begin by reviewing the results of Ref.~\cite{Ahmed2} for the potential $V_4\equiv V_0(1-e^{2|x|/a})$.
By defining $\kappa=\sqrt{V_0-E}$ and $q=\sqrt{V_0}$ the Schr\"odinger equation, in units where $2m/\hbar^2=1$, can be written as
\begin{equation}
\psi''+ (q^2e^{2|x|/a}-\kappa^2)\psi=0.
\end{equation}
This equation can be transformed into the Bessel equation, whose two independent solutions are $J_{\pm \kappa a}(z)$, where $z=qa\,e^{|x|/a}$.
Their behaviour as $x\to\pm\infty$ is governed by the large-$z$ behaviour of $J_\nu(z)$, Eq.~(9.2.2) of Ref.~\cite{AS}, namely $J_\nu(z)\sim\sqrt{2/(\pi z)} \cos(z-\half\nu\pi-\frac{1}{4}\pi)$. That is, they fall off like the square root of an exponential, while oscillating more and more rapidly. Hence they are both square integrable, which is rather surprising from the classical point of view.

In addition to satisfying the Schr\"odinger equation for $x>0$ and $x<0$, the wave-functions $\psi(x)$ must connect smoothly at $x=0$, with no discontinuity in $\psi'(x)$. This is achieved in the combinations
\begin{eqnarray}
\psi_{\mbox{even}}(x)&=&J'_{-\kappa a}(qa)J_{\kappa a}(z)-J'_{\kappa a}(qa)J_{-\kappa a}(z),\nonumber\\
\psi_{\mbox{\small odd}}(x)&=&\mbox{sgn}(x)\left[J_{-\kappa a}(qa)J_{\kappa a}(z)-J_{\kappa a}(qa)J_{-\kappa a}(z).\right]
\end{eqnarray}
The first has $\psi'(0)=0$, the second $\psi(0)=0$.

There is a continuum of such solutions. They are real and square integrable and can be considered as rather unorthodox bound states. However, scattering states can be constructed  by taking appropriate linear combinations of $\psi_{\mbox{\small even}}$ and $\psi_{\rm \small odd}$. Although they fall off rapidly, the probability current is finite because the fall-off in $\psi$ is exactly compensated by the derivative of its rapid oscillations.

In Ref.~\cite{Ahmed2} the authors observed that within this continuum there are special discrete values of $\kappa$, hence $E$, for which the solution $\psi(x)\propto J_{\kappa a}(z)$ alone.  These special values are given by the roots of $J'_{\kappa a}(qa)=0$ for $\psi_{\mbox{\small even}}$ and of $J_{\kappa a}(qa)=0$ for $\psi_{\rm \small odd}$. These occur for $0<E<V_0$. The wave-functions for the two lowest-energy states $\psi_0$ and $\psi_1$ are shown in Figs.~2(a), (b).

In fact, these particular energies turn out\cite{BRath} to be the energy levels of the associated confining potential $V_5=V_0(1-e^{-2|x|/a})$, shown in the lower panel of Fig.~1. The corresponding wave-functions, $J_{\kappa a}(qa\, e^{-|x|/a})$ and $\mbox{sgn}(x)J_{\kappa a}(qa\, e^{-|x|/a})$ are shown in Figs.~3(a), (b) respectively. In this case their behaviour for large $|x|$ is governed by the \underline{small}-$z$ behaviour of the Bessel function, Eq.~(9.1.7) of Ref.~\cite{AS}, namely $J_\nu(z)\sim (\half z)^\nu/\Gamma(\nu+1)$. In this case the eigenvalue condition is that $J_{-\kappa a}(qa\, e^{-|x|/a})$ be excluded, because it blows up at infinity.

\begin{center}
\begin{figure}[h!]
\resizebox{!}{4.5cm}{\includegraphics{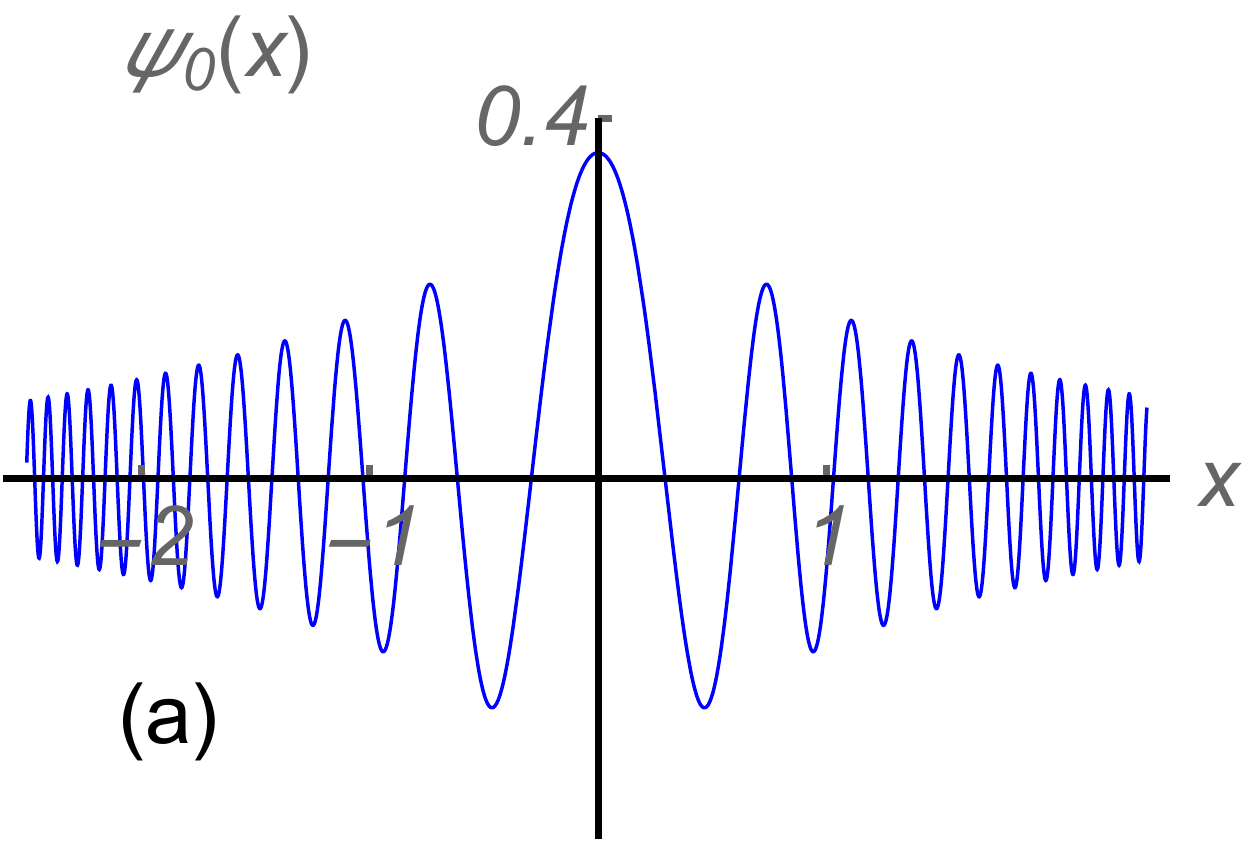}}\hspace{0.5cm}
\resizebox{!}{4.5cm}{\includegraphics{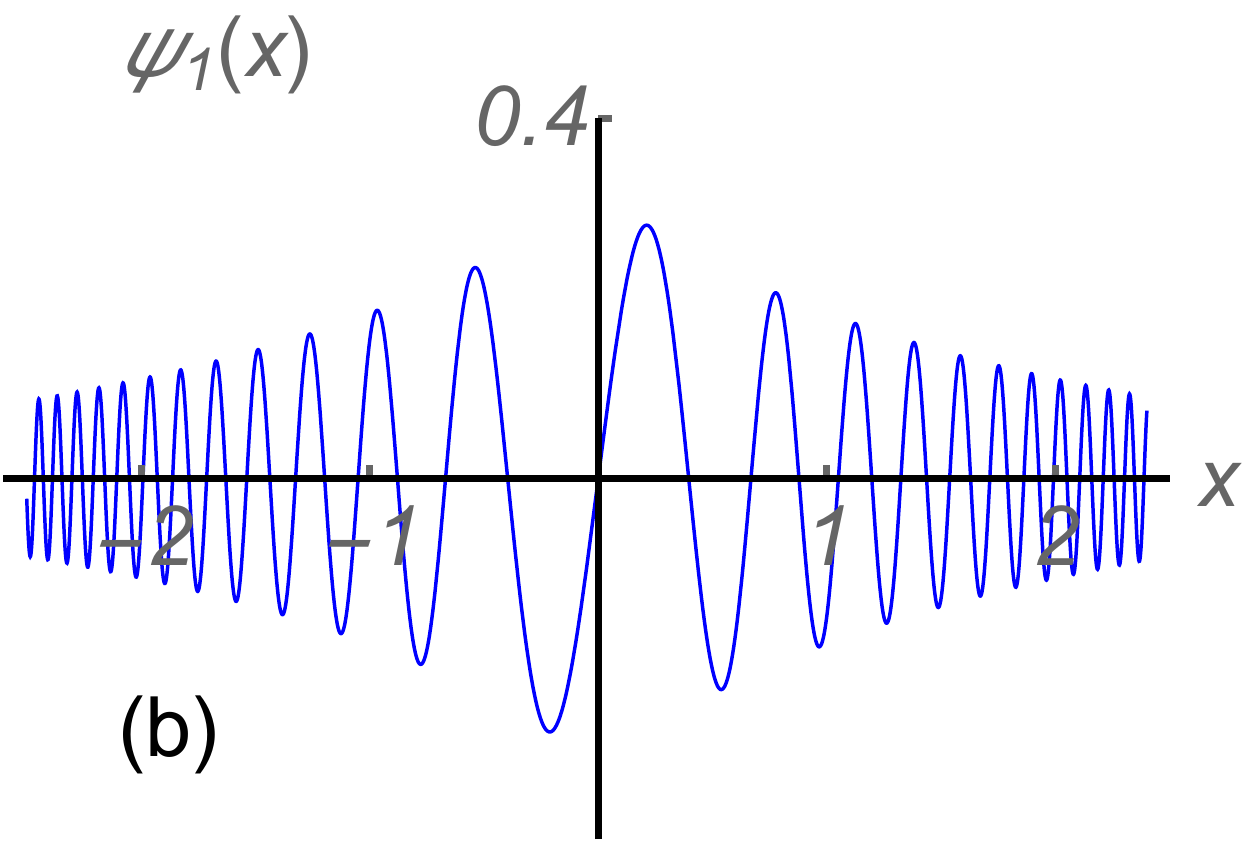}}\hspace{0.5cm}
\caption{Wave-functions $\psi_0(x)$ and $\psi_1(x)$ of $V_4\equiv V_0(1-e^{2|x|/a})$ for the lowest values of $E$ where (a) $J'_{\kappa a}(qa)=0$, (b) $J_{\kappa a}(qa)=0$. In case (b) the sign of the wave-function has been reversed at $x=0$ to avoid a cusp. In this figure and the following two figures we have taken $a=1$, $V_0=50$, in which case $E_0=18.611$ and $E_1=37.263$.}
\vspace{-1cm}
\end{figure}
\end{center}
There are also other discrete energies, given by the roots of $J'_{-\kappa a}(qa)=0$ and $J_{-\kappa a}(qa)=0$, for which the wave functions are proportional instead to $J_{-\kappa a}(qa\, e^{-|x|/a})$, but these seem to have no physical significance.
\begin{center}
\begin{figure}[h!]
\resizebox{!}{4.5cm}{\includegraphics{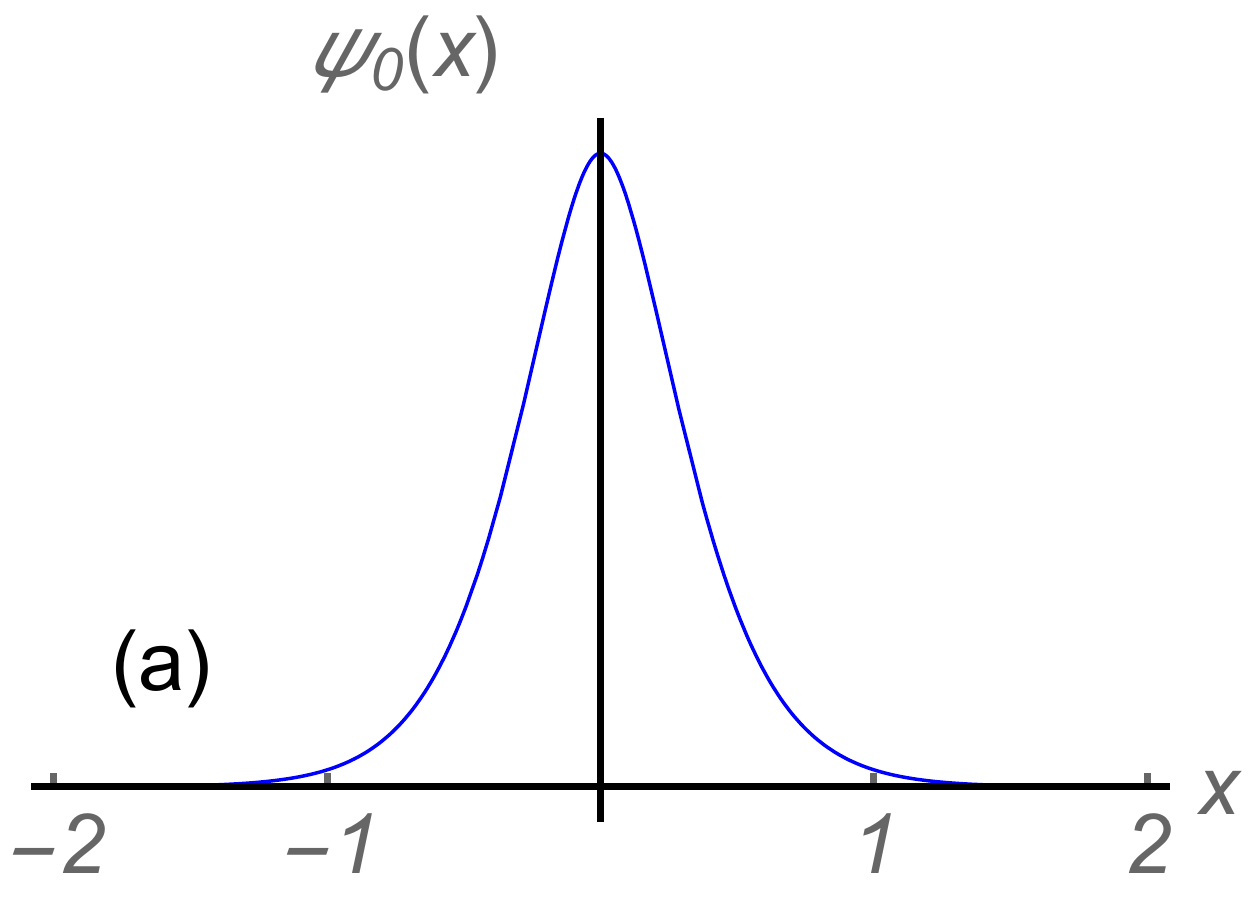}}\hspace{0.5cm}
\resizebox{!}{4.5cm}{\includegraphics{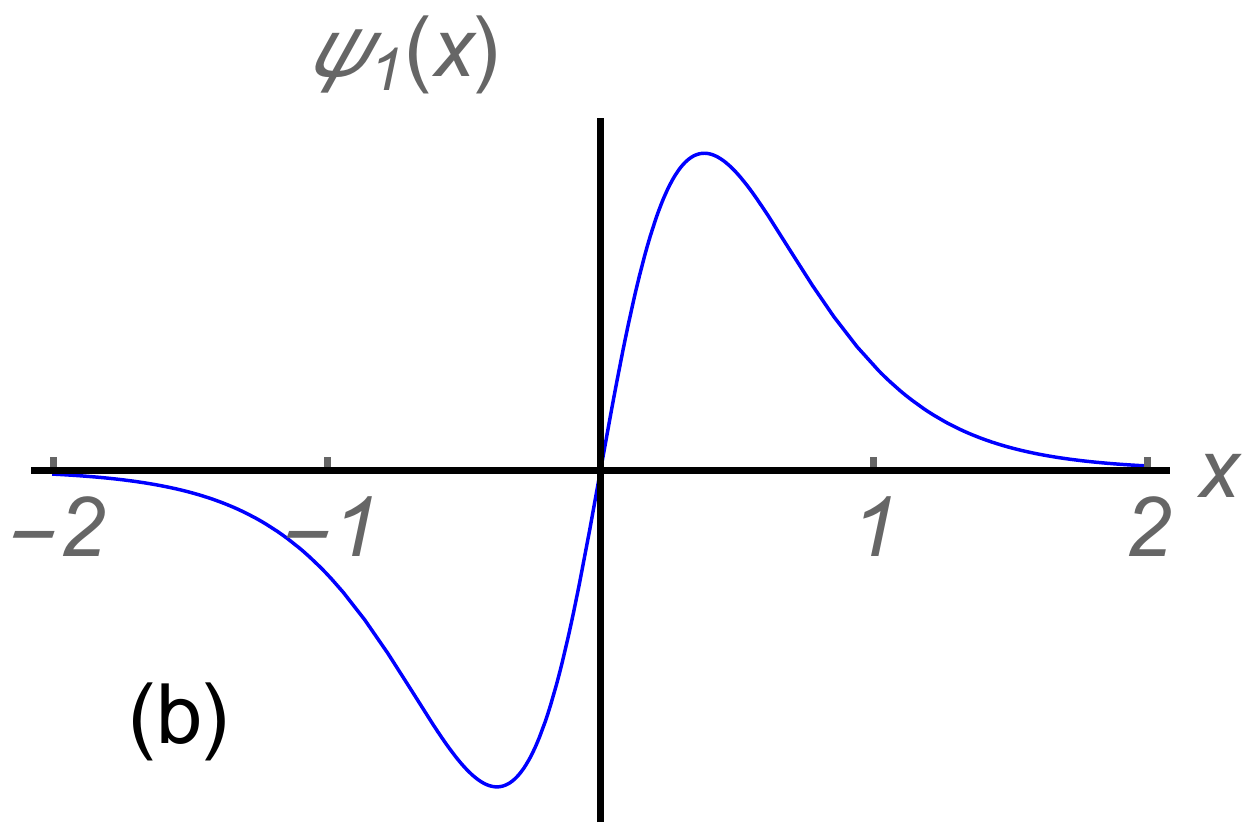}}\hspace{0.5cm}
\caption{Eigenfunctions of $V_5\equiv V_0(1-e^{-2|x|/a})$ for the lowest values of $E$ where (a) $J'_{\kappa a}(qa)=0$, (b) $J_{\kappa a}(qa)=0$. Again the sign of the wave-function has been reversed at $x=0$ in case (b).}
\end{figure}
\end{center}
\vspace{-1cm}
We are now in a position to put these two pieces together by considering the third, analytic, potential $V_6\equiv V_0(1-e^{-2x/a})$ of Fig.~1. This is a smooth continuation from $V_5(x)$ for $x>0$ to the original potential  $V_4(x)$ for $x<0$. Likewise, the eigenfunctions will be those of $V_5$ on the right, matching smoothly onto those of the original potential on the left, as shown in Figs.~4(a), (b).  So we have an analytic potential that goes to $-\infty$ on the left, but nonetheless has genuine discrete eigenvalues and eigenfunctions.
\begin{center}
\begin{figure}[h!]
\resizebox{!}{4.5cm}{\includegraphics{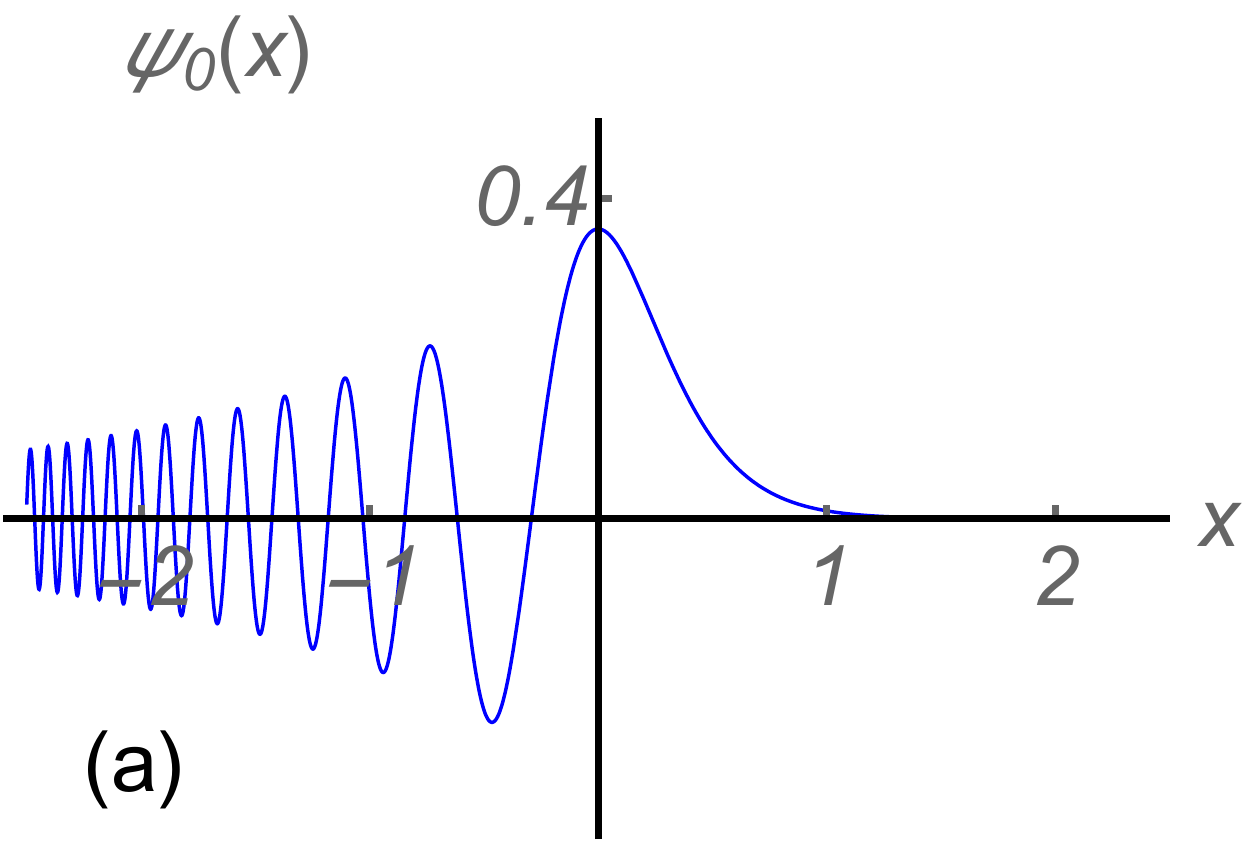}}\hspace{0.5cm}
\resizebox{!}{4.5cm}{\includegraphics{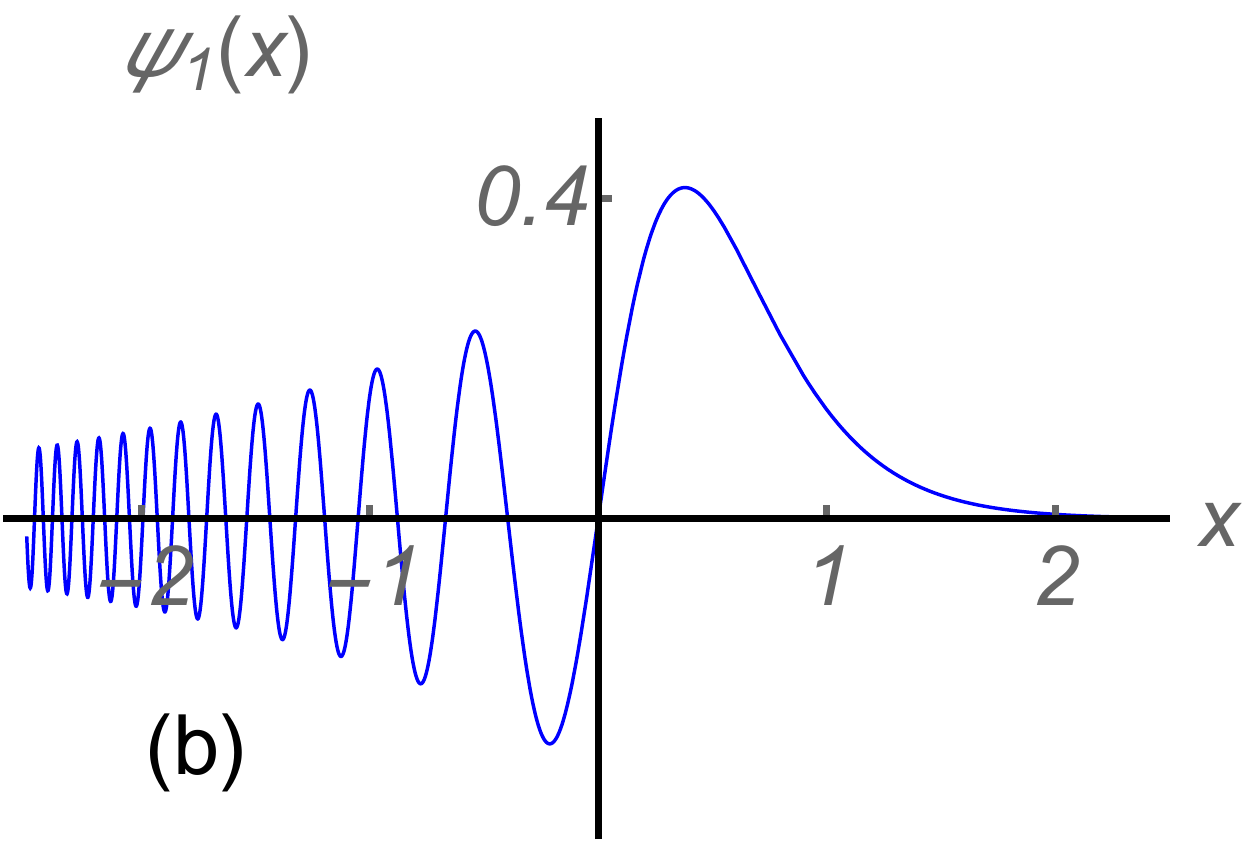}}
\caption{Eigenfunctions of $V_6\equiv V_0(1-e^{-2x/a})$ for the lowest values of $E$ where (a) $J'_{\kappa a}(qa)=0$, (b) $J_{\kappa a}(qa)=0$. }
\end{figure}
\end{center}
\vspace{-1cm}
We will explore the nature and properties of these eigenfunctions in the last section, but now discuss the status of the special discrete energy levels in the context of scattering from the first potential, $V_4$.

Let us consider a wave approaching the potential from the left. It will not be a linear combination of plane waves $e^{\pm i k x}$, but rather of the form $\psi_L(x)= A H^{(2)}_\nu(z) +B H^{(1)}_\nu(z)$, where $H^{(1)}$ and $H^{(2)}=(H^{(1)})^*$ are Hankel functions with index $\nu = \kappa a$ and argument $z= qa e^{-x/a}$. The incoming wave is represented by $H^{(2)}$, and the reflected wave by $H^{(1)}$, because of the asymptotic behaviour $H^{(1)}_\nu(z) \sim \sqrt(2/(\pi|z|)\exp[i(|z|-\half \nu\pi-\quarter\pi)]$. On the right, $\psi=\psi_R(x)= H^{(1)}$, which now represents a right-going wave, because here the argument is $qa e^{+x/a}$. Thus, with  $d/dx=(z/a) d/dz$, the flux is
\begin{equation}
F= \frac{z}{2ia}\ W(H^{(2)}_\nu(z), H^{(1)}_\nu(z))
= \frac{2}{\pi a},
\end{equation}
where $W$ is the Wronskian, given in Eq.~(9.1.17) of Ref.~\cite{AS}. Similarly for the components of $\psi_L$.

Matching the two wave-functions and their derivatives at $x=0$, we obtain
\begin{eqnarray}
H_1&=&A H_2+B H_1,\cr
H_1'&=&-A H_2'-B H_1',
\end{eqnarray}
where we use the shorthand $H_r=H^{(r)}_\nu(z_0)$ ($r=1,2$,\ $z_0= \kappa a$), and the derivatives are with respect to $z$.
The solutions are
\begin{eqnarray}
A&=&- \frac{i\pi z_0}{2}H_1 H_1',\cr
B&=&  \frac{i\pi z_0}{4}(H_1 H_2'+H_2 H_1'),
\end{eqnarray}
where we have again used the value of the Wronskian.

What do the special values of the energy where either $J_\nu(z_0)=0$ or $J_\nu'(z_0)=0$ imply for the scattering amplitudes? When $J_\nu(z_0)=0$ the two Hankel functions are related\cite{AS} by $H_1=-H_2$. Hence
\begin{equation}
B-A=(i\pi z/4)(H_1 H_2'-H_2 H_1') = 1 \hspace{2cm} (J_\nu(z_0)=0).
\end{equation}
 Alternatively, when $J_\nu'(z_0)=0$ the derivatives of the Hankel functions are related by $H'_1=-H'_2$, giving
\begin{equation}
B-A=(i\pi z/4)(H_2 H_1'-H_1 H_2') = -1 \hspace{2cm} (J_\nu'(z_0)=0).
\end{equation}
In fact $B$ is purely imaginary, so writing $B =ib$, we have $A= \pm 1+ib $, with $|A|^2-|B|^2=1$, as required by unitarity, as the transmission and reflection coefficients are given by $T=|1/A|^2$ and $R=|B/A|^2$ respectively.

From these results we can connect with the previous considerations, where we found a static wave-function of the form $J_\nu(z)$. Given that $J_\nu(z)=\half(H^{(1)}_\nu(z)+ H^{(2)}_\nu(z))$, we can regard this as being composed of equal amounts of left- and right-going waves on each side. This situation is described by the transfer matrix $M$, which can be constructed from the above results for scattering from the left and the symmetry properties of $M$. That is, $M_{22}=A$, $M_{21}=-B$, $M_{11}=M_{22}^*$, $M_{12}=M_{21}^*$, giving
\begin{equation}
M=\left(\begin{array}{cc} \pm 1-i b & i b\\ -ib &\pm 1+ib \end{array}\right),
\end{equation}
which has the eigenvector $(1,1)$, as required.

Thus we have connected the scattering properties of $V_4$ to the bound states of $V_5$. A further connection between the different potentials was noted in Ref.~\cite{Ahmed1}, namely that poles in the scattering amplitudes for $V_4$ when the sign of $V_0$ is reversed correspond to the bound-state energies of $V_1$. In that case, $q\to iq$ and $\kappa\to i\kappa$, so that by virtue of the identity\cite{AS} $K_\nu(z_0)= \half i \pi H_\nu^{(1)}(i z_0)$, the Hankel function $H_1$ in the above expression for $A$ goes over to the modified Bessel function $K_{i\kappa a} (qa)$. Then $A=0$ gives either $K'_{i\kappa a} (qa)=0$ or $K_{i\kappa a} (qa)=0$, which, as we will see, are the conditions for the bound states of $V_1$, or the special scattering states of $V_2$, to which we now turn our attention.

For the potential $V_2=-V_0(1-e^{-2|x|/a})$ we define $q=\sqrt{V_0}$ and $\kappa=\sqrt{E+V_0}$, so that the Schr\"odinger equation becomes
\begin{equation}
\psi''(x)-(q^2 e^{-2|x|/a}-\kappa^2)\psi(x)=0,
\end{equation}
whose independent solutions are $I_{i a \kappa}(z)$ and $K_{i a \kappa}(z)$, where $z=qa e^{-|x|/a}$. The special values of $E$ when $\psi$ is proportional to $K_{i a \kappa}(z)$ alone are given by the zeros of $K_{i a \kappa}(qa)$ and $K_{i a \kappa}'(z)$ for odd and even wave-functions respectively.
These occur for $E>0$. The wave-functions for the two lowest-energy states $\psi_0$ and $\psi_1$ are shown in Figs.~5(a), (b).
\begin{center}
\begin{figure}[h!]
\resizebox{!}{4.5cm}{\includegraphics{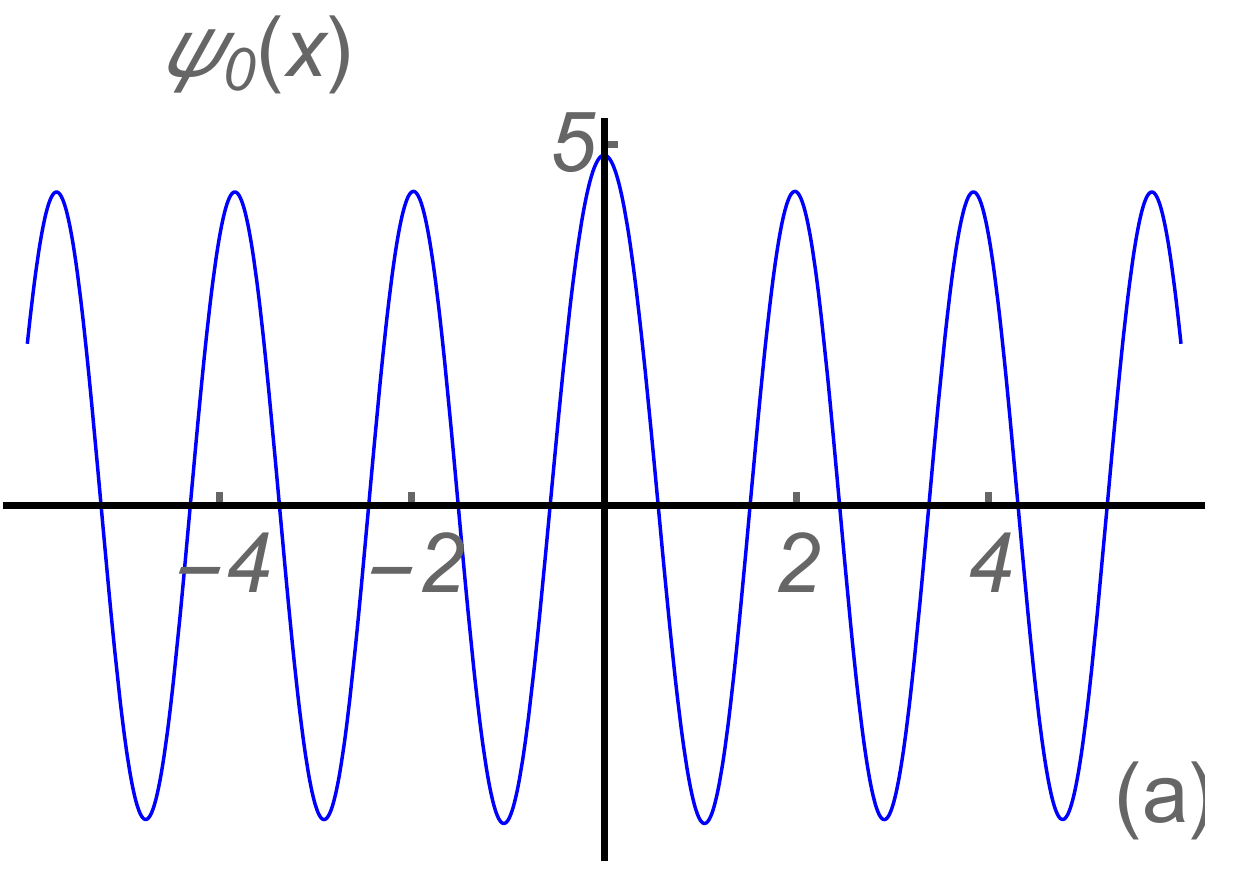}}\hspace{0.5cm}
\resizebox{!}{4.5cm}{\includegraphics{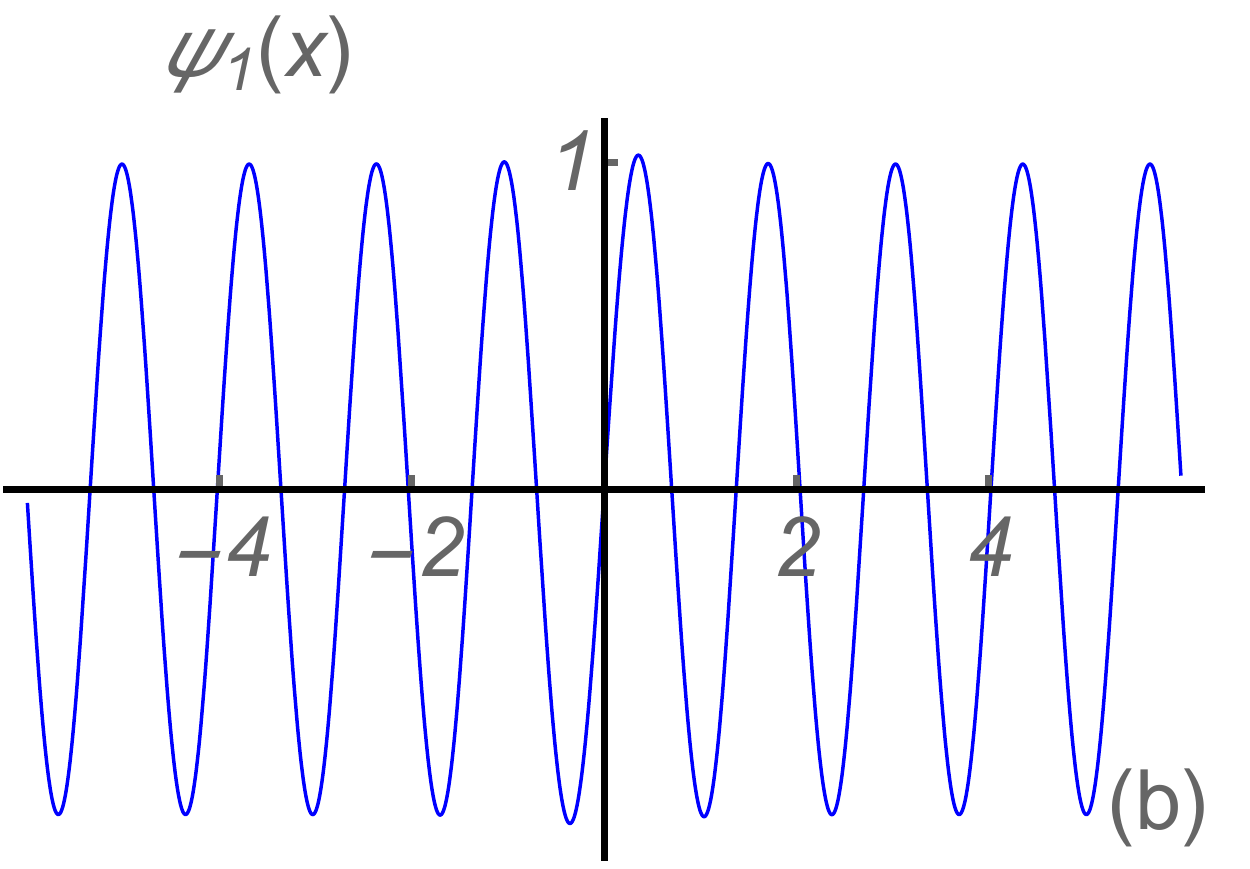}}\hspace{0.5cm}
\caption{Wave-functions $\psi_0(x)$ and $\psi_1(x)$ for the lowest values of $E$ where (a) $K'_{i\kappa a}(qa)=0$, (b) $K_{i\kappa a}(qa)=0$. In case (b) the sign of the wave-function has been reversed at $x=0$ to avoid a cusp. In this figure and the following one we have taken $a=1$, $V_0=5$, in which case $E_0=6.465$ and $E_1=17.537$.}
\end{figure}
\end{center}
\vspace{-1cm}
Similarly to the previous case, the significance of these special values is that they are the eigenvalues of the bound states of $V_1$, for which only the $K$ Bessel functions have the correct behaviour at large $|x|$, and the wave-functions match  smoothly onto the eigenvectors of $V_1$, to give solutions of the analytical potential $V_3$, as shown in Fig.~6.
\begin{center}
\begin{figure}[h!]
\resizebox{!}{4.5cm}{\includegraphics{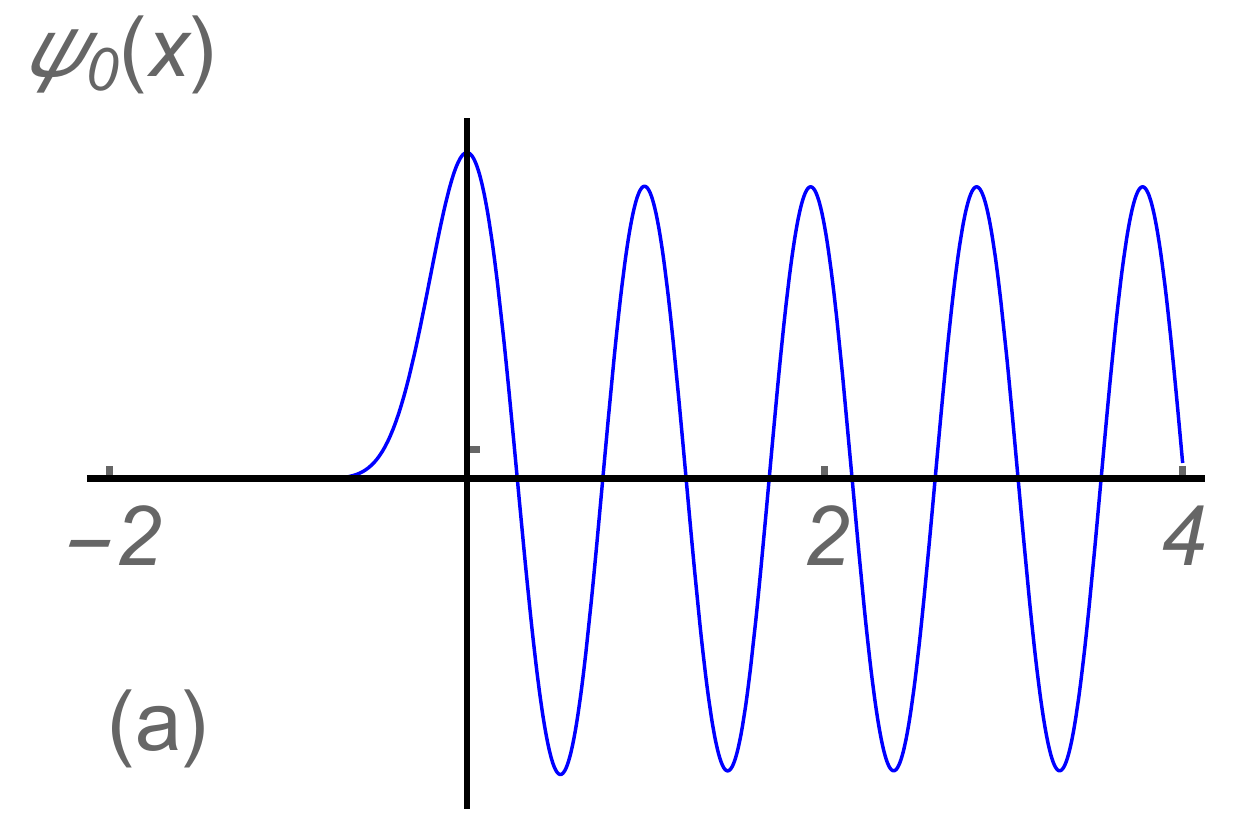}}\hspace{0.5cm}
\resizebox{!}{4.5cm}{\includegraphics{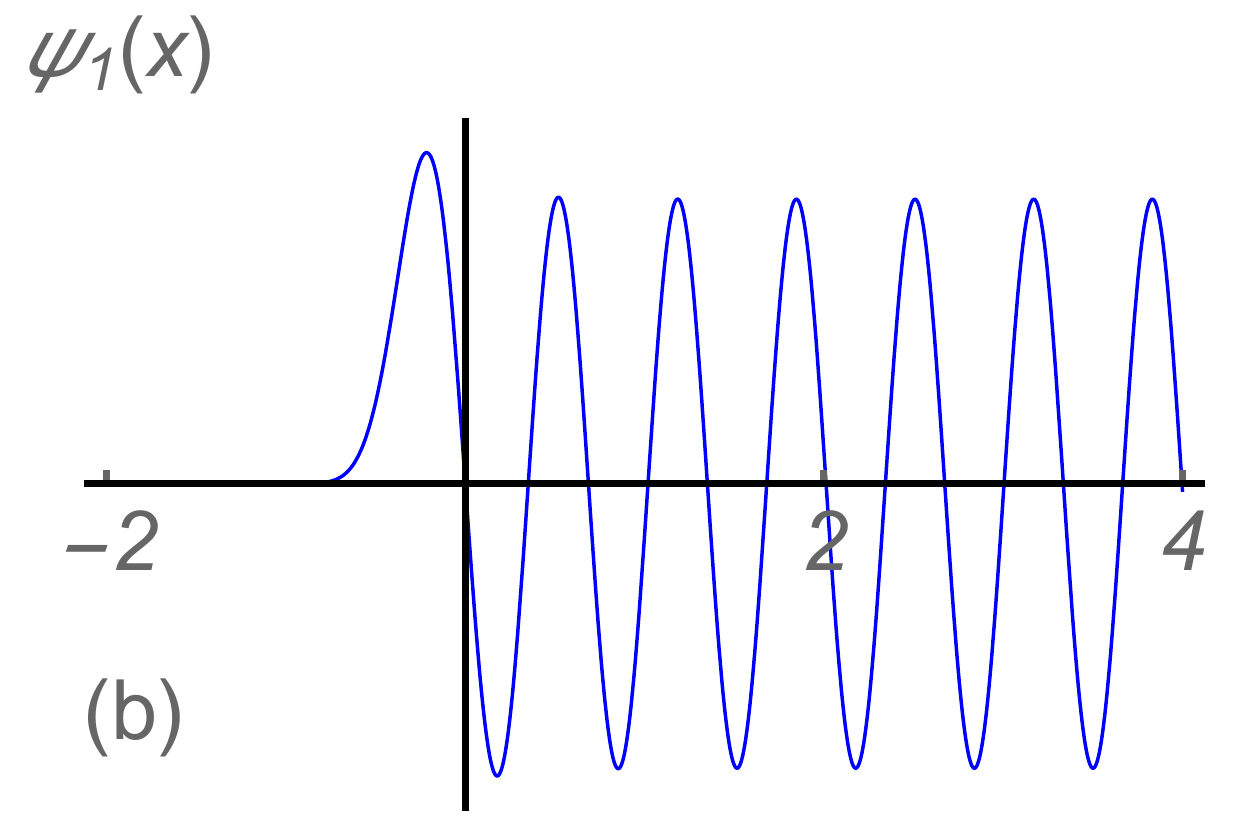}}\hspace{0.5cm}
\caption{ Eigenfunctions of $V_3=-V_0(1-e^{-2x/a})$ for the lowest values of $E$ where (a) $K'_{i\kappa a}(qa)=0$, (b) $K_{i\kappa a}(qa)=0$.}
\end{figure}
\end{center}
\vspace{-1cm}
The final connection is between the potentials $V_2$ and $V_5$ via the scattering amplitudes for $V_2$. Without going into the details we report here the values for the coefficients $A$ and $B$ for scattering from the left, namely
\begin{eqnarray}
A&=&\frac{-i a \pi q}{2 \sinh(a \pi \kappa)}I_{i \kappa a}(qa)I'_{i a \kappa}(qa),\cr&&\cr
B&=&\frac{i a \pi q}{2 \sinh(a \pi \kappa)}(I_{i \kappa a}(qa)I'_{i a \kappa}(qa)+I_{-i \kappa a}(qa)I'_{-i \kappa a}(q a)).
\end{eqnarray}
When $V_0\to -V_0$ the $I$ Bessel functions become $I_{a\kappa}(-iqa)\propto J_{a \kappa}(qa)$ and the poles in the scattering amplitude give the eigenvalues of $E$ for $V_5$.

We have thus established connections between the scattering and bound states of all the different non-analytic pieces of the analytic potentials $\pm V_0(1-e^{2x/a})$.

\section{Discussion}
As has been mentioned above, the wave-functions at the special values of $E$ for $V_4$ or $V_6$ are actually normalizable in spite of the fact that these potentials are both bottomless. They can be considered as static scattering states, or alternatively as rather unorthodox bound states. Because the wave-functions decrease exponentially, all their moments $\langle x^n \rangle$ are finite. However, because of the very rapid oscillations, the expectation values of moments of momenta are not. That is, for $V_4$ every $x$-derivative produces a factor of $dz/dx = \pm q e^{|x|/a}$. The expectation value $\langle p \rangle$ can be taken to be zero, being the integral of an odd function of $x$, but all higher moments diverge.

The lessons we have learned from the exponential potential can be applied to much simpler odd-power potentials $V(x)=x^{2n+1}$, and also non-analytic versions of even power potentials of the form $V(x)=x^{2n-1}|x|$. These all have the property that they rise to $+\infty$ on the $R$ ($x>0)$ and fall to
$-\infty$ on the $L$ ($x<0$). The lowest of these, the linear potential, can be solved analytically, in terms of Airy functions, and the ``odd harmonic oscillator", $V(x)=x|x|$, can be solved in terms of parabolic cylinder functions. Scattering from these potentials is discussed in Refs.~\cite{Bates, Ferreira}. However, they do not possess bound states, because the wave-functions do not fall off fast enough on the $L$. Indeed, their asymptotic behaviour is governed by the equation
$\psi''+|x|^{r-2}\psi=0$, whose solutions are given by $\psi=\sqrt{x}J_{\pm 1/r}(2|x|^r/r)$, with asymptotic behaviour $|x|^{(2-r)/4}$, which requires $r>4$ for convergence. For that reason we concentrate here on the $x^3$ potential ($r=5$), although the picture for higher-order potentials is qualitatively very similar.

In Fig.~7 we show the three associated potentials $V(x)= -|x|^3$, which has decaying, rapidly oscillating wave functions on both $L$ and $R$, $V(x)= +|x|^3$, which has standard bound states, and the analytic potential $V(x)= x^3$ itself, which has hybrid bound states of a similar nature to those of Section I. These are in contrast to the genuine bound states of the $PT$-symmetric potential $V=ix^3$ found by Bender and Boettcher\cite{BB}
\begin{center}
\begin{figure}[h!]
\resizebox{!}{3.3cm}{\includegraphics{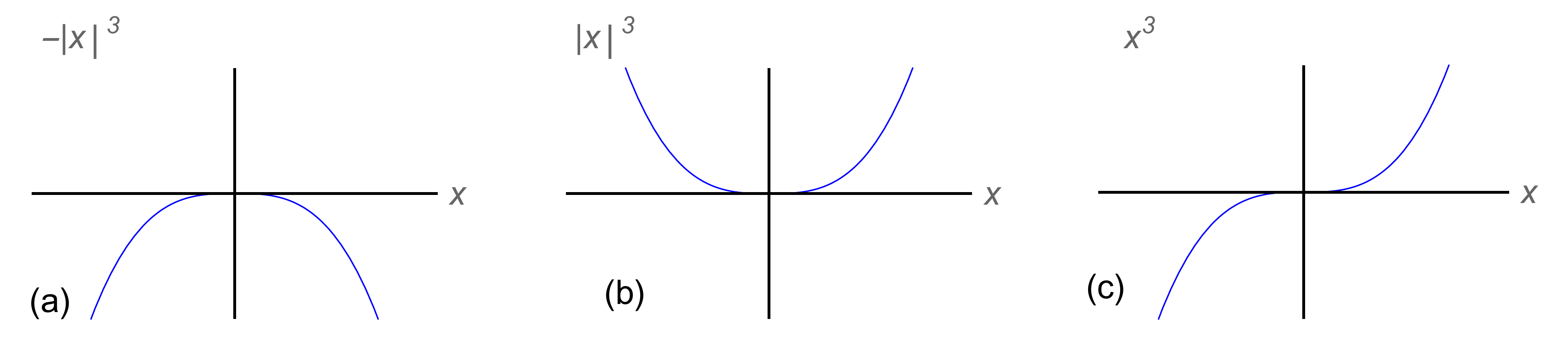}}
\caption{The three associated cubic potentials: (a) $V(x)= -|x|^3$, (b) $V(x)=|x|^3$ and (c) $V(x)=x^3$.}
\end{figure}
\end{center}
\vspace{-1cm}
In this case we do not have an analytic solution, but the eigenvalues and eigenfunctions are readily calculated numerically. The conditions are that $\psi(0)=0$ or $\psi'(0)=0$ and that $\psi(x)$ falls off exponentially for large $x>0$. The first two bound-state eigenfunctions are shown in Fig.~8. The difference here, however, is that the wave-functions are barely normalizable because of the small fall-off for $x<0$, and higher-order moments of both $x$ and $p$ diverge. Hence, although we can establish the relations between the wave-functions of the different segments of the $x^3$ potential, the results are somewhat academic.
\begin{center}
\begin{figure}[h!]
\resizebox{!}{5cm}{\includegraphics{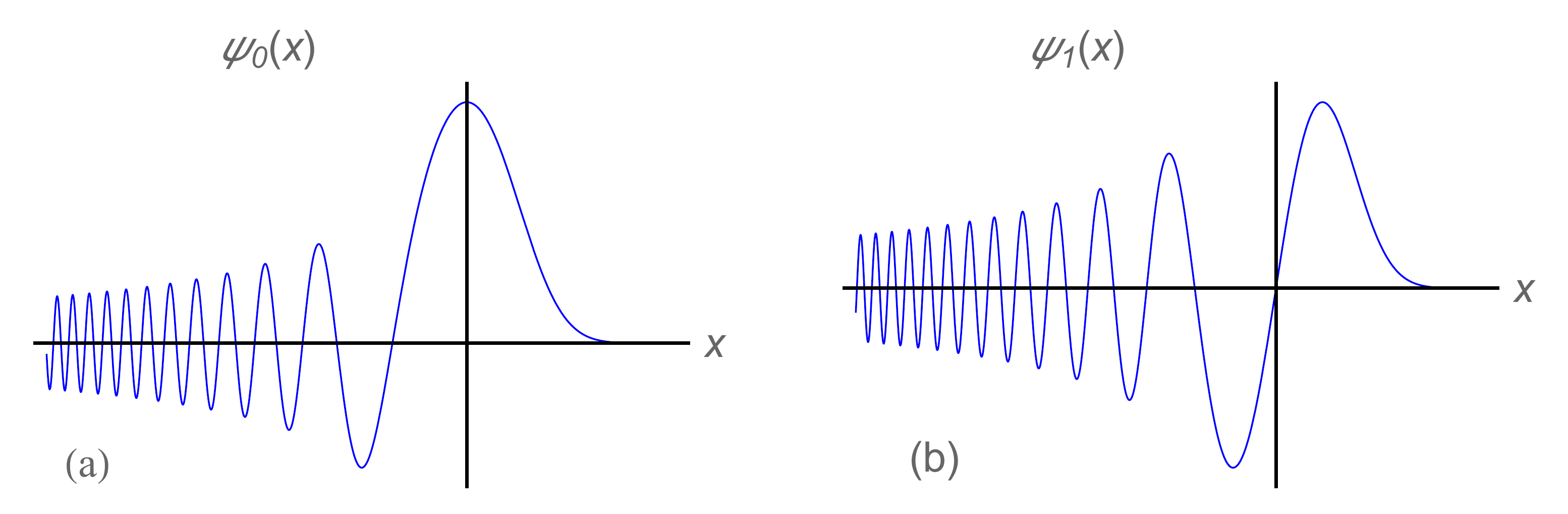}}
\caption{The lowest (a) and first excited state (b) for the potential $V(x)=x^3$. The energies are $E_0=1.023$ and $E_1=3.451$.}
\end{figure}
\end{center}


\end{document}